\documentclass[a4paper, 10pt, conference]{ieeeconf}      
\usepackage{FG2017}
\usepackage{subcaption}
\FGfinalcopy 

\IEEEoverridecommandlockouts                              
\usepackage{graphicx} 

\title{\LARGE \bf
Detecting Cognitive Appraisals from Facial Expressions for Interest Recognition}

\author{\parbox{16cm}{\centering
    {\large Mohammad Soleymani}\\
    {\normalsize
    Swiss Center for Affective Sciences\\
    University of Geneva\\
    Switzerland
    mohammad.soleymani@unige.ch
}}
  \thanks{This work was supported by Swiss National Science Foundation Ambizione grant and in part by Biotech campus foundation}
}
\newcommand{\etal}{\textit{et al.}}

\begin{document}

\ifFGfinal
\thispagestyle{empty}
\pagestyle{empty}
\else
\author{Anonymous FG 2017 submission\\-- DO NOT DISTRIBUTE --\\}
\pagestyle{plain}
\fi
\maketitle

\begin{abstract}
Interest makes one hold her attention on the object of interest. Automatic recognition of interest has numerous applications in human-computer interaction. In this paper, we study the facial expressions associated with interest and its underlying and closely related components, namely, curiosity, coping potential, novelty and complexity. To this end, we conducted an experiment in which participants watched images and micro-videos while a front-facing camera recorded their expressions. After watching each item they self-reported their level of interest, curiosity, coping potential and perceived novelty and complexity. Using an automated method, we tracked facial action units (AU) and studied the relationship between the presence of facial movements with interest and its related components. We then tracked the facial landmarks, e.g., corners of lips, and extracted features from each response. We trained random forests regression models to detect the level of interest, curiosity, and appraisals. We found a large difference between the way people report and react to interesting visual content. The expressions in response to images and micro-videos were not always pronounced depending on the participants. This makes the direct detection of interest from facial expressions a challenging problem. With this work, for the first time, we demonstrate the feasibility of detecting cognitive appraisals from facial expressions which will open the door for appraisal-driven emotion recognition methods.
\end{abstract}

\section{INTRODUCTION}

Interest drives our focus of attention. Recognizing one's interest has broad applications. For example, a recommender system can update its recommendations based on the level of interest shown by the user to a given item. The application of unobtrusive recognition of interest has been explored different domains, e.g., advertisement~\cite{andre_display}, education~\cite{Kapoor2005} and multimedia content summarization~\cite{Gygli2013,soleymani2015quest}. Interest is related to novelty, complexity, comprehensibility and familiarity of the object~\cite{Silvia2008}. Interest is not always related to pleasantness; our attention can be caught by unpleasant events as well, e.g., a car crash scene~\cite{Silvia_book}. 

Research in psychology suggests that interest satisfies the conditions for being an emotion since it has an appraisal structure and bodily expressions~\cite{Silvia_book}. Appraisals are a set of cognitive evaluations in response to an event or object that are important in the construction of emotions~\cite{scherer2005}. For example, when a person looks at an image her mind evaluates the content with regard to its relevance to herself and novelty. Silvia identified the appraisals of coping potential  and novelty-complexity to be the main driving appraisals for interest. The effect of familiarity, comprehensibility and complexity on the interestingness is mediated by the personality of the person~\cite{silvia2009sources}. Emotion, personal connection, familiarity, quality and aesthetics are shown to be important factors in the interestingness of images~\cite{Chu2013,Halonen2011}. 

The existing work on emotion recognition from facial expressions mainly focus on the recognition of prototypical emotions with consistent expressions. Recognition of the continuous emotions in dimensions, such as valence and arousal, has been also explored~\cite{Sariyanidi2015}. There is a limited work on automatic recognition of interest~\cite{BeingBored,Kapoor2005}. To the best of our knowledge, automatic recognition of appraisals have not been reported in the literature.

The past research identified a number facial movements that are active when a person is interested, including eyelid widening and parting lips~\cite{Reeve1993}. Both of these facial action units (AU) are also associated with the appraisal of novelty and coping potential (AU5 and AU 25)~\cite{Scherer2007}. Mortillaro \etal~\cite{Mortillaro2012} proposed that emotion recognition should be done through recognizing cognitive appraisals. If we recognize the appraisals as constructing factors of emotions, we can move beyond the current methods which are mainly based on the automatic recognition of prototypical expressions.   

In this work, we aim at detecting interest, curiosity and appraisals associated with interest, namely, coping potential and novelty-complexity~\cite{Silvia_book}. We first selected a diverse set of images and micro-videos covering a wide range of topics and emotional content. We then recorded a dataset of spontaneous responses from 50 participants watching and rating 80 images and 40 micro-videos. Participants reported their level of interest, curiosity, coping potential (comprehension level), perceived novelty and complexity for each item. A front-facing camera recorded participants' expressions while they were watching and looking at the stimuli. Facial expressions of the participants were analyzed by tracking landmarks and facial action units (AU). We found significant correlations between the present facial action units, appraisals and interest. After registering faces to a standard face, we extracted features from the landmarks in each frame and pooled them for each trial (response to one item) to form a feature vector. We trained an ensemble regression model, i.e., random forests, for detecting appraisals, curiosity and interest from facial expressions. The results demonstrate the ability of facial expressions in capturing patterns associated with appraisals. 
In summary the major contributions of this paper are:
\begin{itemize}
\item For the first time, we report on detecting facial expressions associated with cognitive appraisals.
\item We provide an analysis of the expressions associated with interest and appraisals.
\end{itemize}
In the remaining of this paper, we will familiarize the reader with the existing work, present our material and method, discuss the findings and draw conclusions.
\section{Background}
Silvia \cite{silvia2009sources,Silvia2008} has studied the appraisal mechanism of interest. He found novelty-complexity and coping potential to be the most important appraisals in the process of feeling interest. Coping potential is the ability to cope with an event, for example, in case of images, Silvia used comprehensibility for assessing coping potential~\cite{Silvia2008}. He has also identified that people with a higher level of familiarity with the subject have a higher level of interest in more complex forms of the stimuli. He later found that people can be categorized into different groups regarding how they feel interest towards an object or situation \cite{silvia2009sources}. The first group, with a higher curiosity personality trait, are more likely to be interested by novelty and more complex stimuli. For the second group, however, coping potential and comprehensibility was more important. It is also important to note that interest is not always co-occurring with pleasant emotions and there are unpleasant experiences that might elicit interest \cite{Silvia2008}. In a qualitative study by Halonen \etal~\cite{Halonen2011}, a set of intrinsic characteristics that contribute to the visual interestingness were identified. They included aesthetics, affect, colors, composition, genre, and personal connection.

Reeve \cite{Reeve1993} studied the facial movements as well as physiological responses during the experience of interest. He showed a set of interesting and non-interesting videos to the participants of his experiments. He has identified a set of facial expressions, eye gaze behavior and  head movements, such as head stillness and parting lips, that are associated with interest. Kurdyukova \etal~\cite{andre_display} setup a display that could detect the interest of the passersby by detecting their faces, facial expressions and head pose. Kapoor and his colleagues \cite{Kapoor2004,Kapoor2005} used game state, body posture, facial expressions and head pose to detect interest in children playing an educational game. Body posture was sensed by a grid of pressure sensors installed on the chair where the child was sitting. They could accurately detect interesting situations during the game play. Body posture was the most informative modality for interest detection.  Gatica-Perez \etal~\cite{Simplon2005} proposed a system to recognize the level of interest in a group meeting from audiovisual data.  A dataset of audiovisual recordings from scripted or posed meetings was annotated for the moments of interest, e.g., the moments that people were attentive and took notes. Audiovisual features were extracted from the  participants' faces and voices. The audio channel was the most informative modality in their setting and dataset.  The most comprehensive study on recognition of interest was done by Schuller \etal~\cite{BeingBored}. They recorded an audiovisual interest corpus (AVIC). In their experiment, the experimenter and the participant were sitting on opposite sides of a table. The experimenter played the role of a marketer presenting a product to the participant. The participant was encouraged to engage in a conversation and ask questions. Audiovisual data were recorded and the segmented speaker and subspeaker turns were annotated by the degree of interest on a five points scale.  The five degrees of interest were from disinterest to curiosity. Speech and non-linguistic vocalizations were transcribed and labeled by human transcribers.   Across different modalities,  acoustic features were shown to perform the best.

\section{Material}
\subsection{Stimuli content}
In a preliminary study \cite{soleymani2015quest} a diverse set of 1005 pictures were selected from Flickr\footnote{http://www.flickr.com} covering various topics. Pictures received 20 labels on interestingness, comprehensibility (coping potential), pleasantness, aesthetics arousal, complexity and novelty on Amazon Mechanical Turk\footnote{http://www.mturk.com}. 80 images were selected as stimuli for the current work to cover the whole spectrum in terms of average interestingness, pleasantness and coping potential. 

\begin{figure}
\captionsetup[subfigure]{labelformat=empty}
    \centering
    \begin{subfigure}[b]{0.24\linewidth}
        \includegraphics[width=\textwidth]{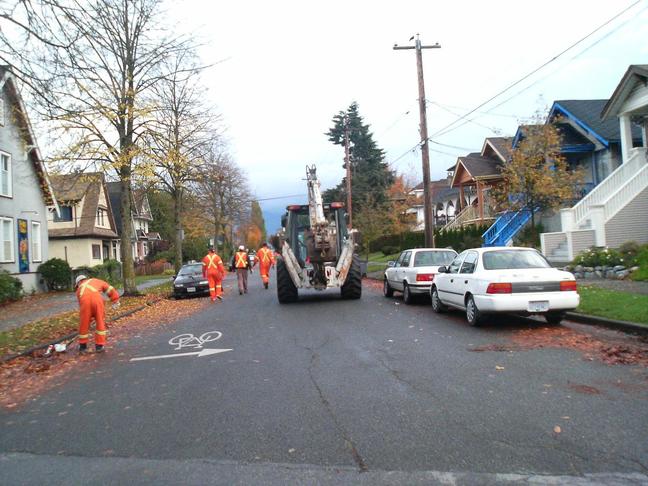}
        \caption{\small Uninteresting}
    \end{subfigure}
    \begin{subfigure}[b]{0.24\linewidth}
        \includegraphics[width=\textwidth]{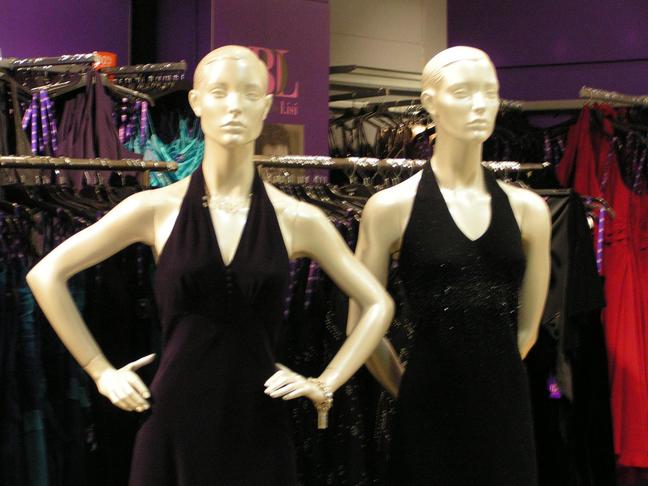}
        \caption{\small Uninteresting}
    \end{subfigure}
    \begin{subfigure}[b]{0.24\linewidth}
        \includegraphics[width=\textwidth]{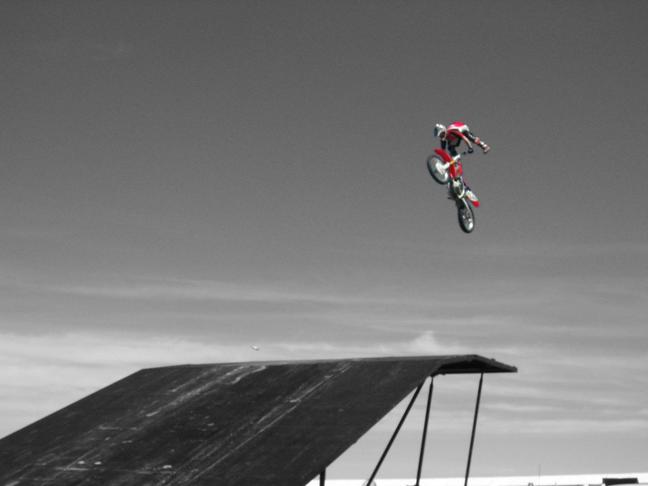}
        \caption{\small Interesting}
    \end{subfigure}
    \begin{subfigure}[b]{0.24\linewidth}
        \includegraphics[width=\textwidth]{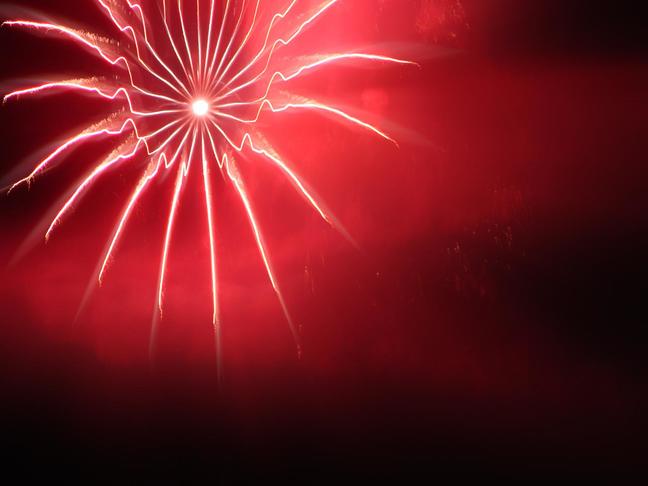}
        \caption{\small Interesting}
    \end{subfigure}
    \\
    \begin{subfigure}[b]{0.24\linewidth}
        \includegraphics[width=\textwidth]{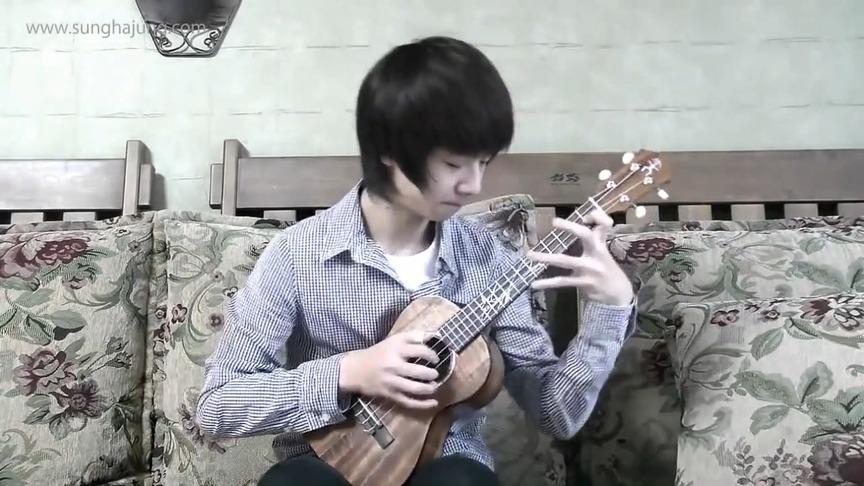}
        \caption{\small Uninteresting}
    \end{subfigure}
    \begin{subfigure}[b]{0.24\linewidth}
        \includegraphics[width=\textwidth]{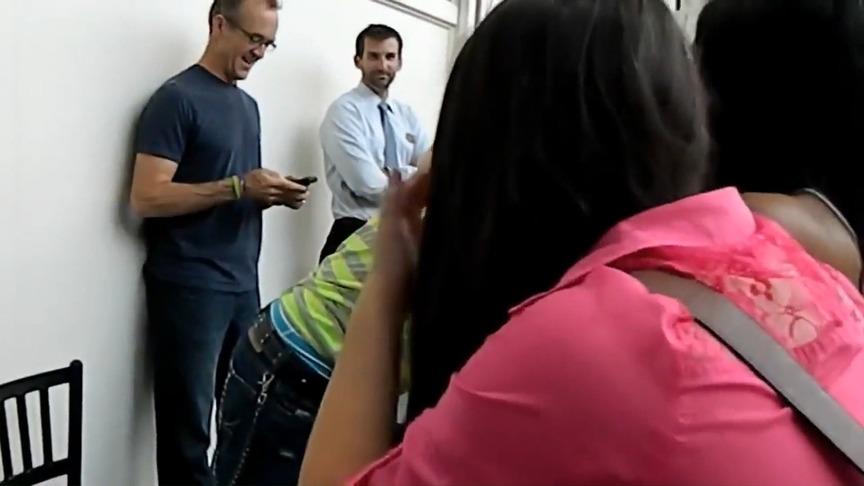}
        \caption{\small Uninteresting}
    \end{subfigure}
    \begin{subfigure}[b]{0.24\linewidth}
        \includegraphics[width=\textwidth]{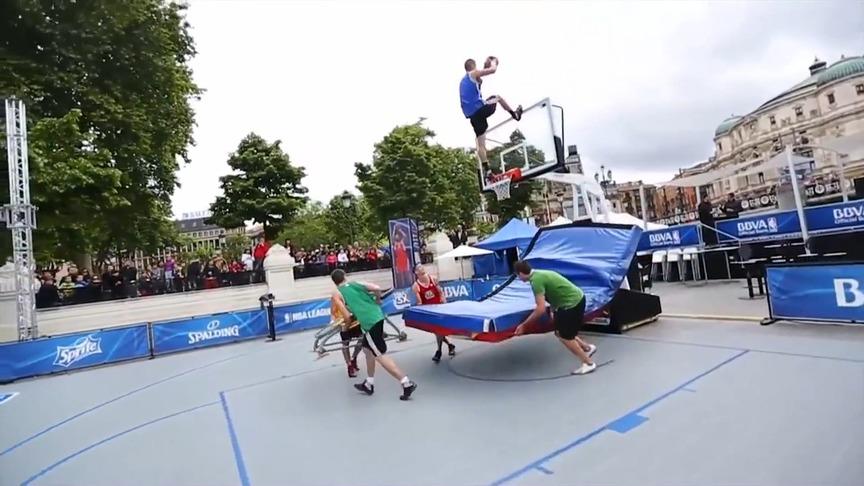}
        \caption{\small Interesting}
    \end{subfigure}
    \begin{subfigure}[b]{0.24\linewidth}
        \includegraphics[width=\textwidth]{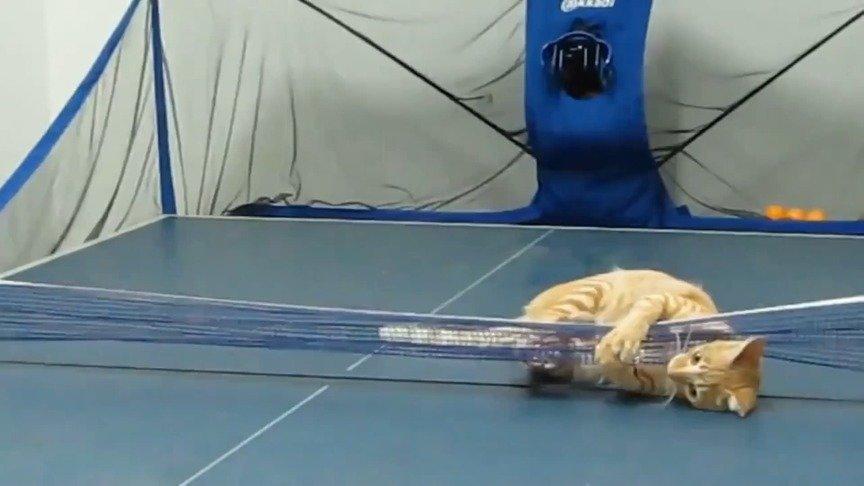}
        \caption{\small Interesting}
    \end{subfigure}
\caption{Examples of stimuli content; the first row shows examples of images and the second row snapshots from micro-videos. Content depicting action and pets are in average more interesting. In micro-videos, content depicting people is often less interesting.}
\label{fig:stimuli}
\end{figure}

132 micro-videos in GIF format from  Video2GIF dataset \cite{vid2gif} were randomly selected and annotated on similar scales on Amazon Mechanical Turk. In our experiments, we displayed the images in full screen mode. GIFs do not have adequate resolution when displayed in a full-screen mode. Hence, we tried to obtain the original versions from the source videos. The Video2GIF GIFs are extracted from YouTube videos. We extracted the higher quality equivalent videos and re-encode them to our desired format with no sound. 40 micro-videos were selected to cover the whole spectrum in terms of average interestingness, pleasantness and coping potential. Since in the experiment we were interested in using the GIF like characteristics of the clips, we re-encoded the videos with 1.5x speed and repeated the sequence twice to demonstrate the possible loopiness. Micro-videos were in average 11 seconds long. Examples of the stimuli are given in Fig. \ref{fig:stimuli}.

\subsection{Recordings}
\begin{figure*}
\captionsetup[subfigure]{labelformat=empty}
    \centering
    \begin{subfigure}[b]{0.20\linewidth}
        \includegraphics[width=\textwidth]{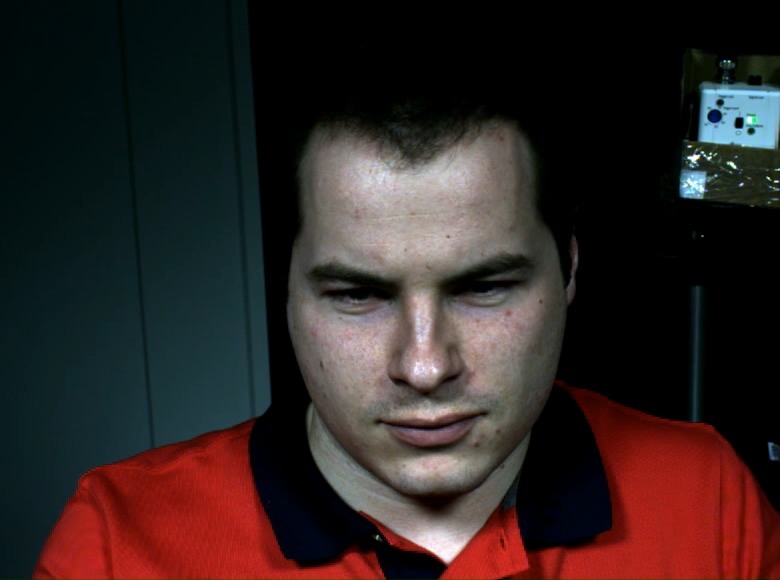}
        \caption{\small Interest}
    \end{subfigure}
    \begin{subfigure}[b]{0.20\linewidth}
        \includegraphics[width=\textwidth]{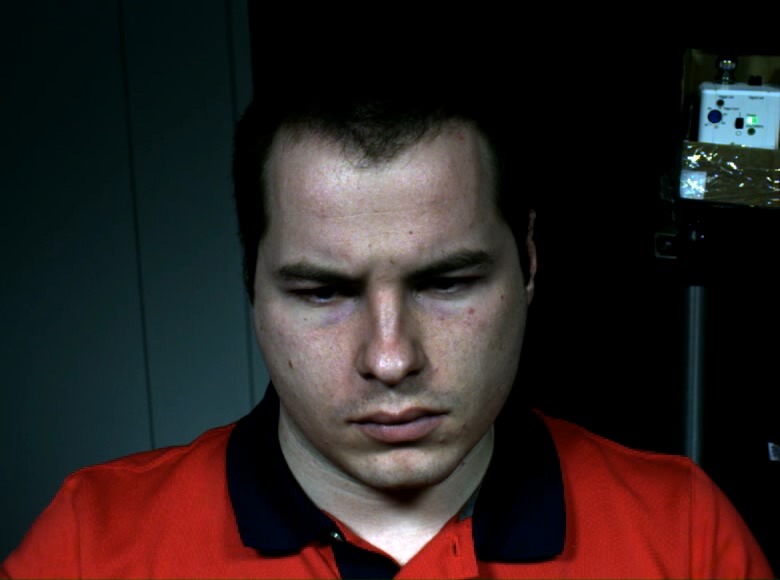}
        \caption{\small Interest}
    \end{subfigure}
	\begin{subfigure}[b]{0.20\linewidth}
        \includegraphics[width=\textwidth]{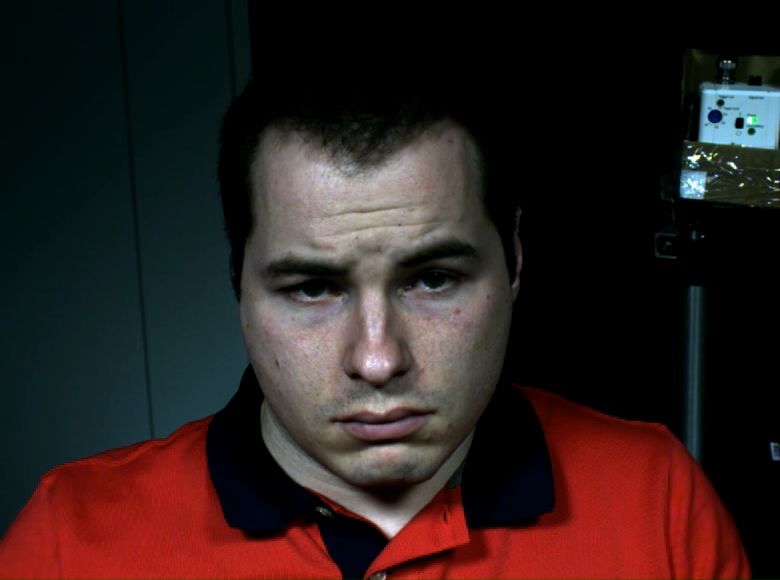}
        \caption{\small Disinterest}
    \end{subfigure}
	\begin{subfigure}[b]{0.20\linewidth}
        \includegraphics[width=\textwidth]{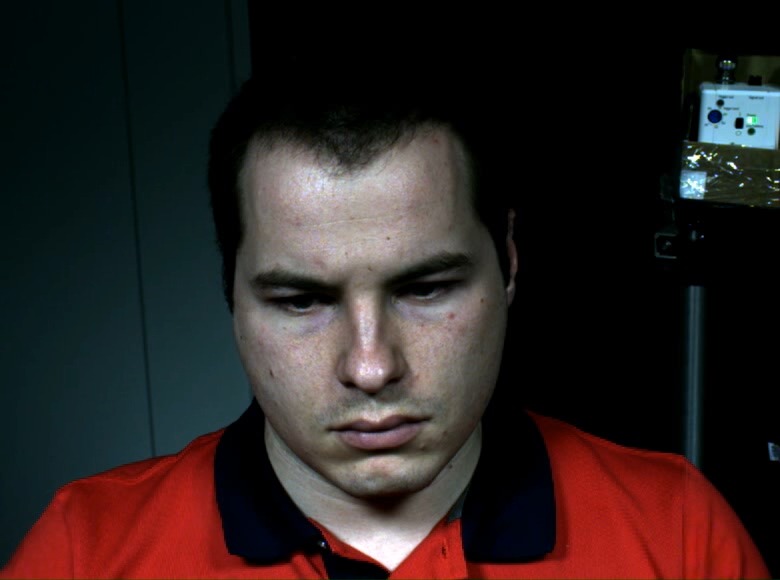}
        \caption{\small Disinterest}
    \end{subfigure}\\
    \begin{subfigure}[b]{0.20\linewidth}
        \includegraphics[width=\textwidth]{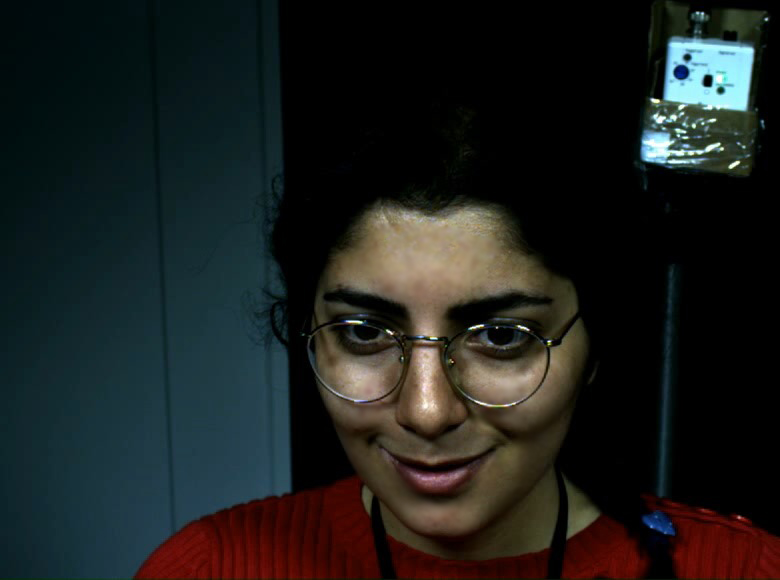}
        \caption{\small Interest}
    \end{subfigure}
    \begin{subfigure}[b]{0.20\linewidth}
        \includegraphics[width=\textwidth]{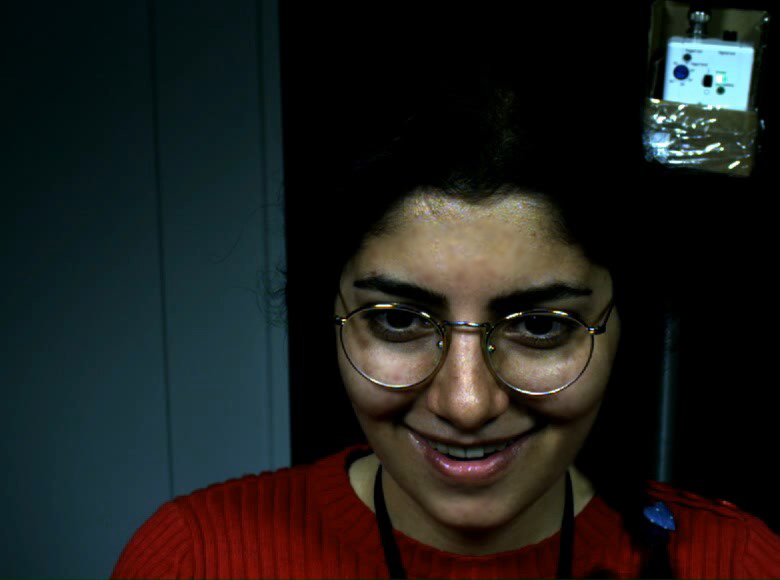}
        \caption{\small Interest}
    \end{subfigure}
	\begin{subfigure}[b]{0.20\linewidth}
        \includegraphics[width=\textwidth]{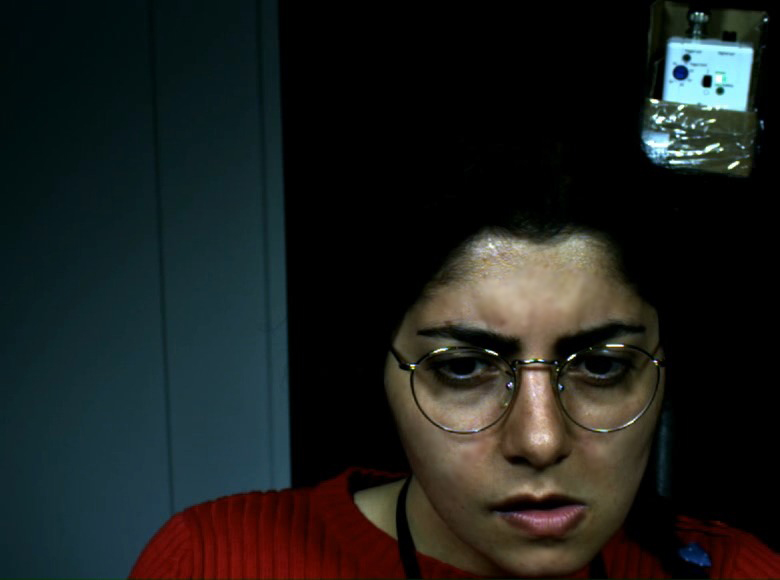}
        \caption{\small Disinterest}
    \end{subfigure}
	\begin{subfigure}[b]{0.20\linewidth}
        \includegraphics[width=\textwidth]{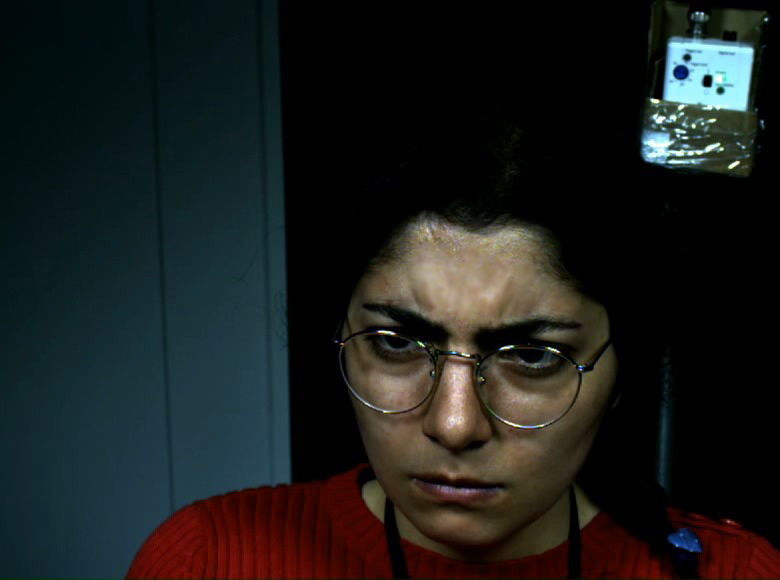}
        \caption{\small Disinterest}
    \end{subfigure}    
\caption{Examples of expressions for interesting ($score=7$) and uninteresting ($score=1$). Participants facial expressions are not consistent. Expressions of the participant in the second row resembles valence more than interest; i.e., for this participant, pleasantness was the main factor for interestingness.}
\label{fig:expr_samples}
\vspace{-10pt}
\end{figure*}
\begin{figure}
\begin{center}
\includegraphics[width=0.6\linewidth,trim={10pt 50pt 10pt 10pt},clip]{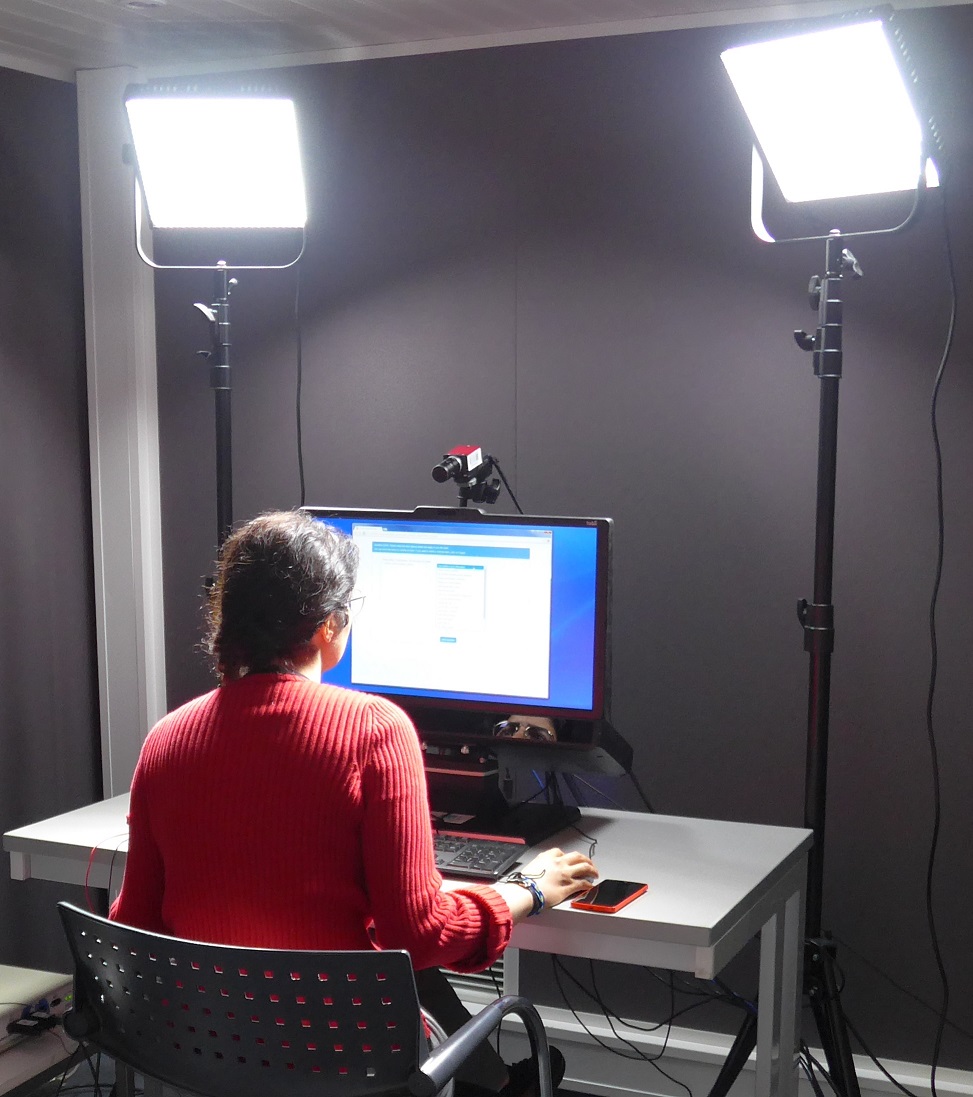}
\caption{The recording setup including an eye gaze tracker, front-facing camera capturing face videos and galvanic skin response.}
\label{fig:setup}
\end{center}
\vspace{-15pt}
\end{figure}

The experiment has received ethical approval from the ethical review board of the faculty of psychology and educational sciences, University of Geneva. 
52 healthy participants with normal or corrected to normal vision were recruited through campus wide posters and Facebook. From these 52 participants, 19 were male and 33 were female. Participants were in average 25.7 years old  ($standard deviation=5.3$). Participants were informed about their rights and the nature of the experiment. They then signed an informed consent form before the recordings. They received a monetary gratitude for their participation. 
Participants were first familiarized with the protocol and ratings, in a dummy run. Experiments were conducted in an acoustically isolated experimental booth with controlled lighting. Video was recorded using an Allied Vision\footnote{https://www.alliedvision.com/} Stingray camera at 60.03 frames/second with 780x580 resolution. Stimuli were presented on a 23 inches screen and participants were seated approximately 60cm from the screen. Two Litepanels\footnote{http://www.litepanels.com} daylight spot LED projectors were used for lighting participants' faces to reduce possible shadows. Video was recorded by Norpix Streampix software\footnote{https://www.norpix.com}. Experimental protocol was ran by Tobii Studio\footnote{http://www.tobii.com/}  and the recordings were synchronized by a sound trigger that marked the frames before each stimulus. To simplify the interface we only provided the participants with a keyboard with numerical buttons that they could use to give ratings (1-7). A picture of the experimental setup is shown in Fig. \ref{fig:setup}. We also recorded eye gaze and galvanic skin responses. In this paper, we only report the analysis on facial expressions. Examples of facial expressions in extreme conditions of interest and disinterest are given in Fig. \ref{fig:expr_samples}.

Participants looked at images for five seconds and rated them on their interestingness, invoked curiosity, perceived coping potential, novelty and complexity on a seven point semantic differential scale. Interestingeness was assessed by rating from uninteresting to interesting. Curiosity was assessed by asking how much they like to watch or look at similar content. Coping potential was assessed by averaging the ratings given to the item on two closely related scales; easy to understand - hard to understand and incomprehensible - comprehensible. Novelty was assessed by rating the items from not novel to novel. Complexity was assessed by rating on simple to complex. The correlation coefficient between different ratings of 120 items (40 micro-videos and 80 images) and their inter-rater agreements are given in Table \ref{tab:corr_alpha}. As expected interest and curiosity have a very high correlation. Coping potential and complexity are also highly correlated which means participants found the more complex sitmuli less comprehensible. Despite the findings of Silvia~\cite{Silvia_book} that coping potential is important in the construction of interest, coping potential does not have any correlation with interest. Inter-rater agreement was calculated by Krippendorff's alpha on ordinal scale. The inter-rater agreements are in the range of slight agreement. The low inter-rater agreements are expected due to the subjectivity of the ratings. 

\begin{table}
  \def\arraystretch{1.3}
  \setlength\tabcolsep{5pt}
  \centering
    \caption{Krippendorff's alpha inter-rater agreement on ordinal scale and Spearman rank correlation coefficient between the ratings.}  
  \begin{tabular}{cccccc}
\textbf{scale}&interest&Coping&Curiosity&Novelty&Complexity\\
Interest  &	- &  0.067&  0.77 &  0.31 & 0.26 \\
Coping    & - &  - 	  &  0.00 &  0.39 & 0.67 \\
Curiosity & - &  -    &  -	  &  0.27 & 0.19 \\
Novelty   & - &  -    &  -    & -     & 0.44 \\
Complexity& - &  -    &  -    & -     &  - 	 \\
\hline\hline
\textbf{Krip. $\alpha$}& 0.22&0.19&0.20&0.17&0.13\\\hline
\end{tabular}
\label{tab:corr_alpha}
\end{table}
\section{Facial expressions of interest and appraisals}
\begin{figure}
\captionsetup[subfigure]{labelformat=empty}
    \centering
    \begin{subfigure}[b]{0.40\linewidth}
        \includegraphics[width=\textwidth]{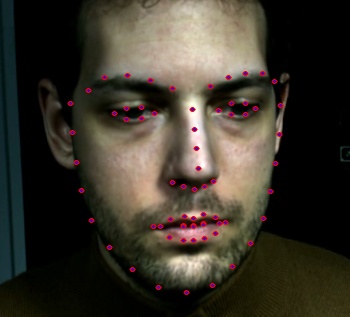}
    \end{subfigure}
    \begin{subfigure}[b]{0.40\linewidth}
        \includegraphics[width=\textwidth]{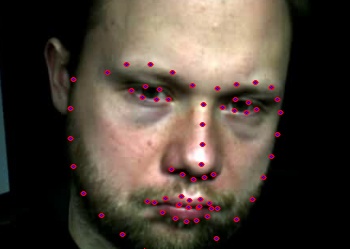}
    \end{subfigure}
\caption{Two examples of cropped detected faces and their tracked facial landmarks overlaid on the original images.}
\label{fig:tracked}
\vspace{-10pt}
\end{figure}
The data from two participants had to be discarded due to the technical failure in recording and synchronization. Head pose, head scale and eye gaze coordinates were extracted in addition to the facial action units. The intensity of the following action units were detected at frame level by OpenFace~\cite{Baltrusaitis2016,Baltrusaitis2015}: AU1, AU2, AU4, AU5, AU6, AU7, AU9, AU10, AU12, AU14, AU15, AU17, AU20, AU23, AU25, AU26 and AU45. OpenFace tracks 68 landmarks on the face (see Fig. \ref{fig:tracked}). After rotating the two-dimensional landmarks from faces to a frontal position and discarding their third dimension, we registered them to a standard face via a rigid transformation calculated by Procrustes analysis on shapes from each frame. We extracted 47 dynamic points on eyes, lips and eyebrows and used their coordinates as features for each frame. The following seven functionals were applied to the features in each trial for pooling: mean, standard deviation, median, maximum, minimum, first and third quartiles. This resulted in a feature vector with 658 elements for each trial. We opted for using landmarks as features since automatic action unit detection has a lower accuracy.

We calculated the Pearson rank correlation between the action units (averaged over each trial) and the ratings. The top three highly correlated action units with each scale are given in Table \ref{tab:corrs}. Interest has the highest correlation with action units associated with smile (AU6 + AU12) in addition to lid tightener (AU7). Curiosity has a similar pattern to interest in addition to AU2 (brow raiser) which is also associated with novelty. Even though, interest is not supposed to be related to pleasantness, the correlation with AU12 shows the bias of our dataset and setting. We hypothesize that both cases of higher levels of pleasantness and unpleasantness are interesting. However, we did not have any extremely unpleasant stimulus in our dataset. Moreover, participants might not self-report higher levels of interest in response to an unpleasant stimuli. Coping potential and complexity are correlated with AU5 which is the eye lid raiser. The more complex or incomprehensible the stimulus, the wider the eyes became. Novelty is surprisingly associated with AU14 (dimpler) and AU23 lip tightener in addition to AU5. The presence of AU14 might be just due to chance or error in AU detection. It is also worth noting that the software we used does not have a high accuracy in detecting all action units \cite{Baltrusaitis2015} and these results are not comparable with the studies in psychology with manual action unit coding.

We were also interested in analyzing whether the presence of expression by itself is a sign for interest or not. All the videos were continuously annotated by one annotator using a software similar to FEELTRACE~\cite{cowie2000feeltrace}. The ratings were given to the amount of expressions present in every frame which we called intensity. We averaged the continuous annotation for every trial and called it the expression intensity score. We calculated the Spearman rank correlation between the intensity score and the ratings given by the participants. The correlation coefficients are given in the last column of Table \ref{tab:corrs}. Interest and curiosity are correlated with the intensity of the expressions, albeit too weakly to assign strong associations. Coping potential and complexity both have almost no correlation. Novelty has the highest correlation among the scales with the intensity of expressions. 

\begin{table}
  \def\arraystretch{1.3}
  \setlength\tabcolsep{5pt}
  \centering
    \caption{Top three most correlated action units (AU) and five scales. $\rho$: Spearman rank correlation coefficient. Only correlation coefficients superior to 0.05 are included ($p<0.0001$). The last column shows the correlation between the average score given to the intensity of expressions with different scales (Intens.: intensity; $^*$ implies significance $p<0.0001$). }  
  \begin{tabular}{ccccccc|c}
\textbf{scale}&AU&$\rho$&AU&$\rho$&AU&$\rho$&Intens.\\\hline
Interest  &	AU12 & 0.099 &AU6  &0.071  &AU7  &0.062&0.065$^*$\\
Curiosity & AU12 & 0.112 & AU6  & 0.074  & AU2 &0.070&0.064$^*$\\
Coping    &  AU5& 0.093&-  & -  &-  &-&0.00\\
Novelty   &  AU23&0.119 & AU5 & 0.105 &AU14  &0.088&0.118$^*$\\
Complexity& AU5 & 0.113 &- & - &- &-&0.025\\
\hline
\end{tabular}
\vspace{-10pt}
\label{tab:corrs}
\end{table}
\section{Experimental results}
\subsection{Appraisal and interest detection}
We used an ensemble regression model, random forests, with 200 trees and minimum leaf size of five for detecting the level of interest, curiosity and appraisals. The strength of such an ensemble method is its lower susceptibility to over-fitting. In our preliminary experiments, random forests outperformed Support Vector Regression with a Radial Basis Function kernel. Due to the ordinal nature of the scores, we opted for rank-normalization for labels from each participant. In rank-normalization, first all the values are sorted and then the rankings will be converted to values between zero and one. Features were normalized by subtracting their mean and dividing by their standard deviation. We used 20-folding cross-validation strategy for evaluating the regression results on five different scales, namely, interest, curiosity, coping potential, novelty and complexity. We also performed the regression on each participants' responses using a leave-one-out cross-validation. Results were evaluated using Spearman rank correlation coefficients, due to the ordinal nature of the scores. We also report root-mean-square error and concordance correlation coefficient (CCC), for reference. The regression evaluation results are given in Table \ref{tab:prediction}.

We observed that participants often kept a still face till the stimulus was over and became expressive afterwards. We hypothesize that they were trying to remember the image or video for the ratings and the expressions right after the end of the stimulus could be associated with recall. We also analyzed the expressions of extended excerpts (face video during the time that the stimulus was present in addition to three following seconds). The results of the analysis on the extended trials are given in Table~\ref{tab:predictionplus3sec}.

As expected, results on interest and curiosity are very similar. For inter-participant cross-validation, coping potential and complexity were better detected compared to interest. 

The intra-participant results are inferior compared to the inter-participant ones. This is partly due to the lower number of training data in that case. Overall, coping potential and complexity were detected with higher accuracy compared to interest. This can be associated to the differences between the expressions of interest, e.g., smile and eyes open. It appears that the expressions of appraisals such as perceived complexity is more consistent between participants. This patterns can be also observed for the intra-participants results, except in the case of the extended responses where interest and curiosity are in average better detected.  

Overall, the results of the extended trials were superior to the exact trials, accordance with our observations.  Nevertheless, the results from the extended excerpts follow a similar pattern.

\begin{table}
  \def\arraystretch{1.5}
  \centering
    \caption{Regression results on interest, curiosity, coping potential, complexity and novelty. For RMSE the maximum value is 1. For intra-participant (per participant) results, median and standard deviation values are given.}  
  \begin{tabular}{cccc}
\textbf{Scale} &\textbf{Spearman $\rho$ $\uparrow$} & \textbf{CCC $\uparrow$} &\textbf{RMSE}$\downarrow$ \\
    \hline
    \multicolumn{4}{c}{Inter-participant}\\
    \hline
  Interest &0.25&0.14&0.32 \\
  Curiosity & 0.28&0.15&0.32\\
  Coping potential & 0.45&0.25&0.24\\
  Novelty & 0.35&0.22&0.33\\
  Complexity& 0.42&0.24&0.28\\\hline   
    \multicolumn{4}{c}{Intra-participant}  \\  
     \hline
  Interest &0.10(0.16)&0.11(0.16)&0.31(0.08) \\
  Curiosity & 0.09(0.15)&0.05(0.10)&0.33(0.07)\\
  Coping potential & 0.10(0.12)&0.04(0.08)&0.27(0.04)\\
  Novelty & 0.16(0.14)&0.10(0.08)&0.33(0.06)\\
  Complexity& 0.08(0.11)&0.04(0.07)&0.29(0.05)\\\hline
    \end{tabular}
    \label{tab:prediction}
\end{table}

\begin{table}
  \def\arraystretch{1.5}
  \centering
    \caption{Results from the extended expressions in which facial expressions in three seconds after the end of stimulus were also analyzed. For RMSE the maximum value is 1. For intra-participant (per participant) results, median and standard deviation values are given.}  
  \begin{tabular}{cccc}
\textbf{Scale} &\textbf{Spearman $\rho$ $\uparrow$} & \textbf{CCC $\uparrow$} &\textbf{RMSE}$\downarrow$ \\
    \hline
    \multicolumn{4}{c}{Inter-participant}\\
    \hline
  Interest &0.32&0.21&0.31 \\
  Curiosity & 0.33&0.20&0.31\\
  Coping potential & 0.47&0.28&0.24\\
  Novelty & 0.36&0.24&0.33\\
  Complexity& 0.45&0.29&0.28\\\hline   
    \multicolumn{4}{c}{Intra-participant}  \\  
     \hline
  Interest &0.19(0.19)&0.11(0.16)&0.31(0.07) \\
  Curiosity & 0.16(0.18)&0.10(0.14)&0.32(0.07)\\
  Coping potential & 0.12(0.14)&0.07(0.11)&0.27(0.04)\\
  Novelty & 0.09(0.17)&0.05(0.12)&0.34(0.06)\\
  Complexity& 0.12(0.14)&0.06(0.10)&0.28(0.05)\\\hline
    \end{tabular}
    \label{tab:predictionplus3sec}
    \vspace{-10pt}
\end{table}


\subsection{Discussions}
In our results, the detection accuracy varied widely by participant. There are two sources of inconsistencies across participants; first, they did not interpret interest in the same way; second, the expressions also varied both in terms of their intensity and pattern. As it is evident in the examples given in Fig. \ref{fig:expr_samples}, some participants' expressions of interest and disinterest was very similar to valence or pleasantness. There were also a number of participants who did not show much visible expression. To test the hypothesis of whether the general expressiveness of the participant was related to the performance of appraisal and interest detection, we calculated the correlation coefficient between average expressiveness score of each participant and the average performance (measured by correlation) for interest and appraisal detection. We found significant correlation between the expressiveness score and coping potential ($r=0.35$, $p=0.01$), curiosity ($r=0.45$, $p=0.001$) and complexity ($r=0.45$, $p=0.001$). The  correlation demonstrates that the method performs worse for less-expressive participants. Essentially, we are only able to use similar methods for expressive people. Hence, facial expression as a single modality is unable to accurately capture the level of interest for less-expressive people. 

The existing work on the automatic recognition of interest did not find facial expressions to be the most informative modality~\cite{BeingBored,Simplon2005} which demonstrates the challenging nature of recognizing interest only from facial expressions. For interest, unlike Ekman basic emotions~\cite{ekman1993facial}, there is no evidence that there is a unique and consistent facial expression. As an alternative, recognizing appraisals can be combined with the content analysis to detect interest from both expressions and the content. For example, if intrinsic pleasantness and aesthetics are related to interest in images, as is shown in [removed for double blind review] 
the visual content can be analyzed or tagged on the degree of its pleasantness and aesthetics. The recognized appraisals, such as novelty, can be then used with intrinsic pleasantness for interest detection.

In this work, the content was limited to images and micro-videos with no personal connection to the participants. However, in practice the relevance of the content or personal connection to the user is an important factor in determining its interestingness. A grainy picture of a loved one might be more interesting than a sharp and aesthetically pleasing image of a random scene. This limitation can be addressed in the future by adding personally relevant and irrelevant content to assess the appraisal of relevance. The other limitation of this work is that the participants did not have any specific task or goal. The more passive a person, it is less likely that they feel or express any emotion.

Our results are not at the same level as the ones reported by \cite{BeingBored}. However, there are a number of differences in the experiment and analysis. First, the protocol in \cite{BeingBored} consist in an active social interaction whereas our recordings were done in non-social setting where participants are less expressive. Second, their ground-truth was generated by the third-person labelers which are more consistent compared to the self-reports with participant-dependent bias. 
\section{Conclusions}
In this work, we conducted an experiment with the goal of assessing visual interest and its related components. Analysis on the action units showed that interest is related to eye opening and smile, which are signs of novelty and pleasantness. The correlation between smile and interest is in agreement with our previous findings in a similar context which showed positive correlation between pleasantness and interest~\cite{soleymani2015quest}. We found the problem of recognizing visual interest in a passive setting challenging due to three problems. First, there is no easy and consistent way for self-reporting interest. Second, some people are not very expressive in such a context. Third, there is no unique expression of interest. 

From the appraisals, novelty was the one with the most pronounced and consistent facial action units. Both coping potential and complexity were only associated with lid raiser and wider eyes. Regarding the detection results, in average, the appraisals were better detected compared to interest itself. The results demonstrated the limitations of facial expressions in detecting interest from facial expressions in a single observation. In the future, we will analyze other modalities to perform multimodal analysis for the task at hand.

\section{ACKNOWLEDGMENTS}
We would like to thank Danny Dukes for discussions and Michael Gygli for his help with extracting Video2GIF original videos.

\bibliographystyle{IEEEtran}
\bibliography{references}
\end{document}